\documentclass[superscriptaddress,showpacs,aps,prb,floatfix, preprint]{revtex4-2}
\usepackage{bm,amsmath,amssymb}
\usepackage{graphicx}
\newcommand{\breite}{0.8} %% set 0.8 for one column or 1 for two column style

\begin{document}
\title{Capacitance and conductance oscillations from electron tunneling into high energy levels of a quantum well in a p-i-n diode}

\author{S. Parolo}
\author{M. Lupatini}
\author{E. Külah}
\author{ C. Reichl}
\affiliation{Solid State Physics Laboratory, ETH Z\"urich, CH-8093 Z\"urich, Switzerland}
\author{W. Dietsche}
\affiliation{Solid State Physics Laboratory, ETH Z\"urich, CH-8093 Z\"urich, Switzerland}
\affiliation{Max-Planck-Institut f\"ur Festk\"orperforschung, D-70569 Stuttgart, Germany}
\author{ W. Wegscheider}
\affiliation{Solid State Physics Laboratory, ETH Z\"urich, CH-8093 Z\"urich, Switzerland}

\begin{abstract}

Two-dimensional  electron and hole gases separated by a few nm from each other are produced in p-i-n diodes based upon MBE-grown GaAs/AlGaAs heterostructures. At such interlayer distances, the exciton formation and possibly Bose-Einstein condensation (BEC) is expected. We measure the capacitance between the layers and find it to oscillate as a function of the bias voltage. The peak values exceed the geometric capacitance by up to a factor of two. The surprisingly regular periods of the oscillations are of the order of 10 to 30 mV and scale linearly with the inverse of the thickness, between 60 and 150 nm, of the GaAs layer placed between the barrier and the p-doped AlGaAs. The oscillations are related to the resonant electron tunneling into high energy levels of this GaAs layer acting as a quantum well. We suggest that long lifetimes of the electrons in these levels are the origin of the large peak values of the capacitance. The possible relation of the capacitance oscillations with BEC is discussed.
\end{abstract}

\maketitle

\section{Introduction}

The Bose-Einstein condensation (BEC) of excitons in semiconductors continues to be one of the most interesting subjects in solid state physics. Excitons are quasi-particles formed by interacting electrons and holes with an integer total spin.\cite{Blatt1962a,Lozovik1975,Shev1976} Earlier experiments using optically excited electrons and holes did not yield conclusive evidence about BEC because the intrinsically rapid recombination of the excitons limits their coherence and condensation.\cite{Snoke2002}
A considerable enhancement of the lifetime is achieved by separating the electrons and holes in spatially separated quantum wells combined with an applied electric field.\cite{Butov:2002bh,Beian2017}
%In the case of "dark" excitons the recombination rate is further reduced by a spin mismatch to the photons and signatures  of BEC have been observed.\cite{PhysRevLett.126.067404}
%An alternative route to long lifetimes is offered by exciton-polaritons where excitons are strongly coupled to light in an optical resonator leading to quasi-particles showing clear signatures of BEC.\cite{Deng02,Kasprzak:2006ul}
An alternative route is offered by exciton-polaritons where excitons are strongly pumped by light in an optical resonator leading to quasi-particles showing clear signatures of BEC.\cite{Deng02,Kasprzak:2006ul}
 Quasi-excitons with infinitely long lifetimes form in electron or hole bilayers with two half-filled Landau levels in a quantizing magnetic field.\cite{Wie2004,Eis2004,Tut2004} Josephson currents and dissipationless flow of the exciton condensate has been demonstrated.\cite{Spi2000,Tie2008,Xuting2012} 
Quasi-excitons exist also in p-i-n devices with a barrier as demonstrated with GaAs based devices\cite{Sivan1992,Pohlt2002,Croxall2008,Seamons2009} as well as in 2D atomic layer structures\cite{Wang:2019fk,Burg_2018,L_pez_R_os_2018}.

In this paper, we report on experiments using p-i-n structures in the GaAs/AlGaAs system with a 10-nm barrier and measure its capacitance rather than lateral transport. We also place a quantum well with thicknesses between 60 and 150 nm directly adjacent to the barrier. The capacitance oscillates with increasing forward voltage reaching peak values twice the geometric capacitance. We suggest that resonant tunneling into the quantum well generates long lived exciton states leading to the enhanced capacitances.

 \begin{figure}
	\centering
	\includegraphics[width=\breite\columnwidth]{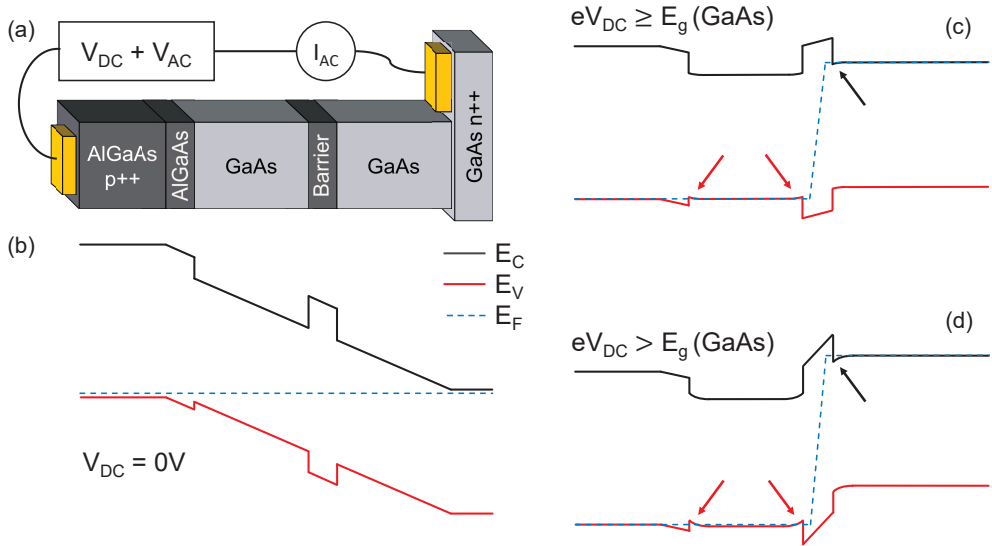}
	\caption{\textit{(a) Layer sequence of a p-i-n structure with a central barrier and setup to measure differential capacitance and conductance. (b) to (d) schematic band alignments at three different bias voltages. Only the $\Gamma$ bands of the conduction electrons and the heavy-holes valence bands are shown.
			 The quantum well on the p-side of the barrier does not significantly change its shape from (c) to (d). The black and red arrows mark the locations where 2DEGs and 2DHGs, respectively, form.}}
	\label{fig:LayersAndBandstructures}
\end{figure}

\section{Experiment}

\textit{\textbf{Samples:}} We use molecular-beam epitaxy (MBE) to grow  p-i-n devices with a typical layer sequence as shown in Fig.~\ref{fig:LayersAndBandstructures}(a). More details and other modifications of the structure are described in the supplement.\cite{Sup} An n-doped GaAs layer is first grown on a semi-insulating GaAs substrate, followed by an undoped GaAs spacer layer. A 10-nm barrier consisting of $\text{Al}_{0.8}\text{Ga}_{0.2}\text{As}$ is then grown as a short period superlattice.\cite{Pohlt2002}  For comparison we also produced samples with a 20-nm barrier. The respective barrier is followed by another undoped GaAs spacer with 60 to 150 nm thickness. The sample is completed with a 30 nm thick undoped $\text{Al}_{0.3}\text{Ga}_{0.7}\text{As}$ layer and a 100 nm  thick p-doped  $\text{Al}_{0.3}\text{Ga}_{0.7}\text{As}$ contact layer. Positive DC voltages are applied to the p-contact of the device and lead to both a 2D electron and a 2D hole gas (2DEG and 2DHG) at the barrier as soon as the voltage exceeds the value corresponding to the GaAs band gap energy.

   Schematic band structures of such devices are shown in Fig.~\ref{fig:LayersAndBandstructures}(b)-(d) for different values of the voltage.  At zero voltage, the area between the doped layers is depleted. At an applied bias corresponding to the GaAs gap energy, a 2DEG on the n-side of the barrier forms as well as a 2DHG on the p-side. Actually, there are now two 2DHGs on the p-side, the one directly at the barrier with a density increasing with voltage and another one at the GaAs/AlGaAs boundary, with a density that varies only slightly with bias. Thus, the band structure on the p-side is reminiscent of a quantum well with holes at its boundaries and a nearly flat band bottom that keeps its shape also at higher voltages,  Fig.~\ref{fig:LayersAndBandstructures}(d).

 The mobilities of the two-dimensional charge layers in our devices cannot be measured directly, but electron mobilities of more than 30 million $\text{cm}^2/\text{Vs} $ have been achieved in the same growth campaign. Thus, our samples should have a very low impurity level. The actual samples are produced by standard lithography and measure about 230 x 240 $\mu \text{m} ^{2} $. Contacts to the bottom n-doped layer are made by AuGeNi and to the top p-layer by TiPtAu. 

\begin{figure}
	\centering
	\includegraphics[width=\breite\columnwidth]{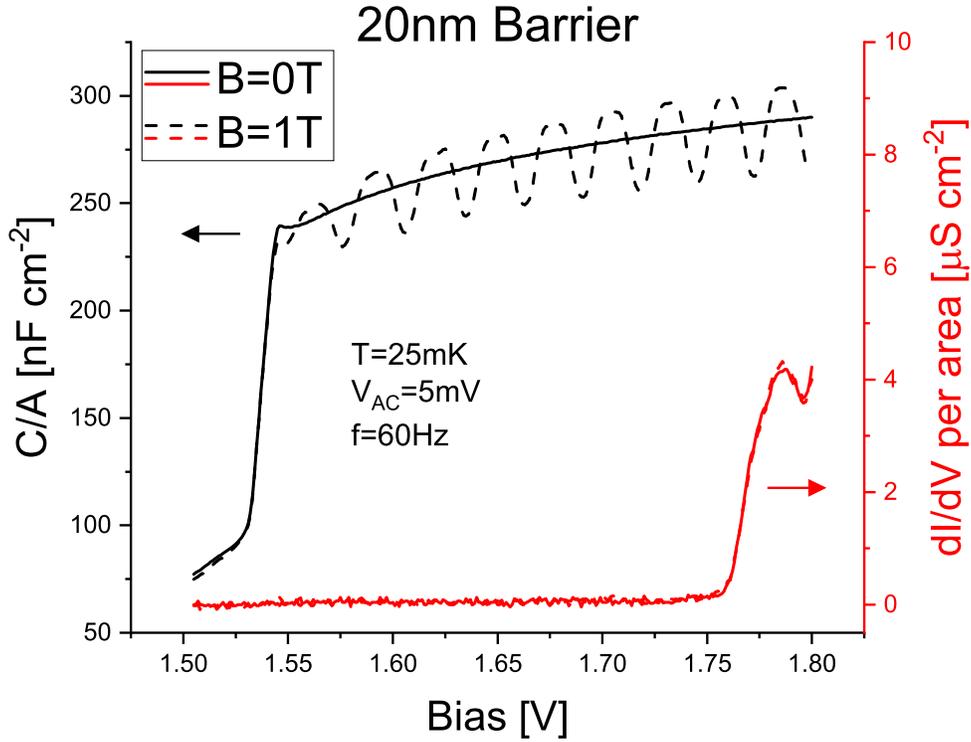}
	\caption{{\small \textit{Differential capacitance measured with a 20-nm barrier sample in zero magnetic field and with a 1 T magnetic field, respectively. The two-dimensional charge gases form above 1.53 V. In magnetic field the capacitance shows a series of minima corresponding to  the minima of the density of states between the Landau levels. The ohmic conductivity (red trace) from tunneling or current across defects is less than the noise level up to about 1.75 V.}}}
	\label{fig:20nmLandauOscillations}
\end{figure}

\textit{\textbf{Experimental techniques}} The differential conductance between the p- and the n-layer is measured by applying a DC bias with a small added AC voltage. Both the in-phase and the out-of-phase component of the resulting AC current are measured with a current amplifier and a lock-in. The out-of phase current is proportional to the device capacitance  and increases linearly with frequency. The in-phase component of the current measures the ohmic contribution due to tunneling or leakage via defects in the barrier. This is equivalent to a parallel circuit of a capacitor and a resistor.
The interpretation of our data depends on constant Fermi energies over the p and n-side of the barrier, respectively. This has already been assumed for the band-structure schematics of Fig.~\ref{fig:LayersAndBandstructures}(b)-(d).

\textit{\textbf{20-nm barrier}}. We first present results on a sample with a 20 nm barrier in order to verify our measurement technique. The sample is of the type (a) in Fig. SM.1 in the supplement \cite{Sup} with 60 nm thick GaAs spacers. Tunnelling is weak with such a wide barrier. We find that the values of the capacitance deduced from the out-of-phase currents do not vary with frequencies between 10 Hz and 3 kHz by more than a few percent. This indicates that charges on the p- and the n-side of the barrier equilibrate on a time scale of less than one millisecond.
 Data of the capacitance and of the differential ohmic (in-phase) conductance, both normalized to an area of one cm$^2$, are shown in Fig.~\ref{fig:20nmLandauOscillations}. 
  Note that the capacitive contribution of the leads, typically 25 pF, has been determined separately and is already subtracted for the plot.  The capacitance signal shows a steep increase at 1.53 V, which is very close to the band gap energy of GaAs. At this voltage, the energy bands are flat and the 2D charge gases start to form. The capacitance  $ C $ above the step continues to increase with voltage because the extent of the wave functions of the two charge layers shrink with the increasing band bending. It is also modified by interaction effects\cite{Eisenstein1994} leading to a quantum capacitance  $C_{Q}=e^{2}D_{T}$  contribution according to $1/C=1/C_{geo}+1/C_Q$ where $ C_{geo} $ is the geometric capacitance and $ D_{T} $ the thermodynamic density of states.

Applying a perpendicular magnetic field leads to the Landau quantization of the densities of states and therefore to a modification of the quantum capacity. Landau oscillations, here measured at 1 T,  are clearly resolved in the data of Fig.~\ref{fig:20nmLandauOscillations} and are shown as the dashed line. The observation of Landau oscillations verifies that two-dimensional charge systems exist. At 1 T spin splitting is not yet resolved. We note that the Landau oscillations from holes and electrons fall on each other for the supposed equal densities of electrons and holes. The density varies between zero and about 4$ \cdot $10$ ^{11} \text{cm}^{-2}$ from 1.53 to 1.75 V.

 The value of $ C_{geo} $, the geometrical capacitance between the 2DEG and the 2DHG, can be determined from the Landau periodicity because the minima of the capacitance occur whenever the density of electrons/holes increases by $ B/(h/e) $, or 2 times this value if spin is not resolved. From the average Landau periods of Fig.~\ref{fig:20nmLandauOscillations} we find an average $ C_{geo}$=280 $\text{nF}/\text{cm}^2$ corresponding to an average distance d=36 nm   between the 2DEG and the 2DHG. This is close to an average distance of 31 nm found by simulations with the program \textit{Nextnano} \cite{Nextnano2007}. The capacitance calculated from the Landau oscillations is close to the ones measured at zero field and in the Landau maxima. Measurements in fields up to 12 T (not shown) continue to show the Landau oscillations including the spin and fractional gaps and are comparable with those reported by Khrapai et al.\cite{Khrapai2008}. Fig.~\ref{fig:20nmLandauOscillations} shows also the conductivity derived from the in-phase AC current across the device. It is less than the noise (0.2 $\mu S/cm^2$)  up to 1.75 V, indicating that there is no leakage current. Beyond, an ohmic conductivity of a few $ \mu \text{S}/\text{cm}^2$ appears, which is independent from the AC frequency, and is probably due to hot electrons crossing the barrier.

\begin{figure}
	\centering
	\includegraphics[width=\breite\columnwidth]{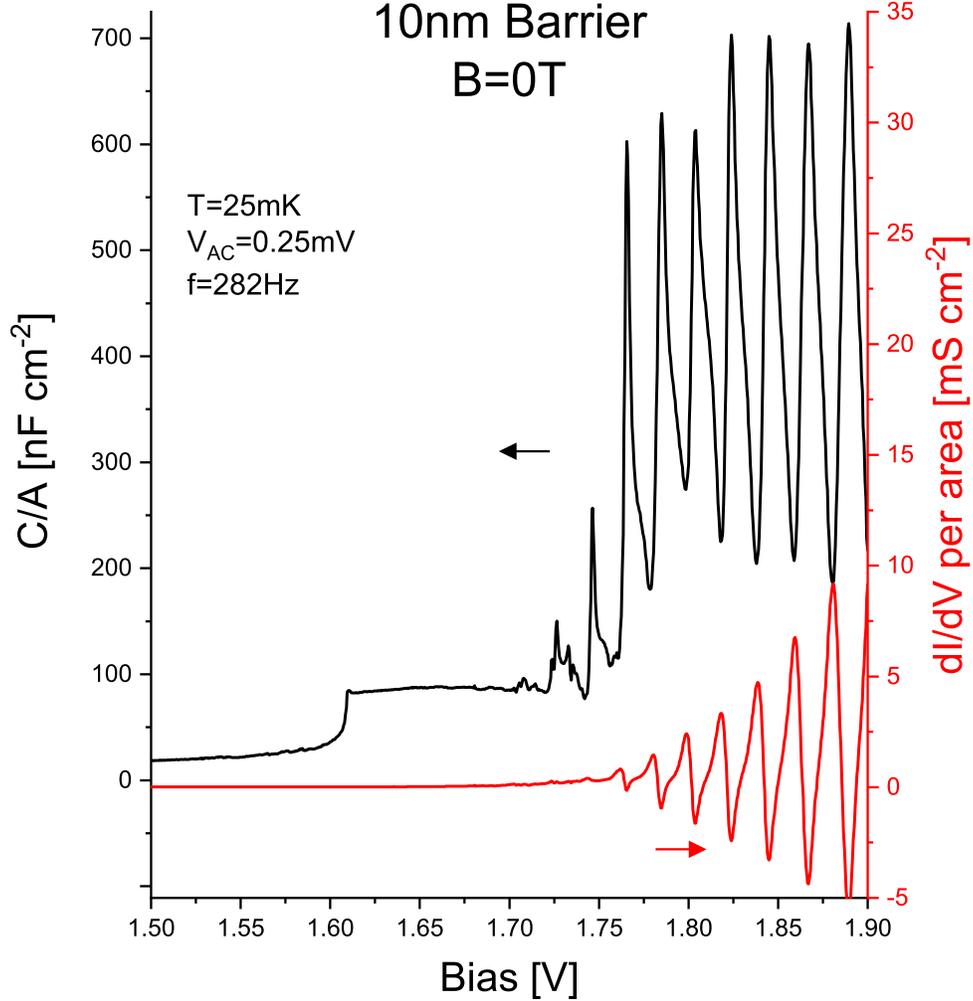}
	\caption{{\small \textit{Capacitance and ohmic conductance in a sample with a 10 nm barrier and a 90 nm wide well on the p-side measured at 25 mK. At bias voltages exceeding about 1.7 V,  regular capacitance oscillations with a large amplitude are observed.}}}
	\label{fig:10nmZeroFieldOscillations}
\end{figure}

\textit{\textbf{ 10-nm barrier}} At 10 nm, the distance between the positive and negative charges is small enough to show BEC phenomena as demonstrated with bilayers at half-filled Landau levels \cite{DingXuting2014}. In the experiments presented here, we find capacitance oscillations without any magnetic field applied, of which the origin is unclear.
 Plots of both the differential capacitance and of the ohmic conductance are shown in Fig.~\ref{fig:10nmZeroFieldOscillations} measured at about 25 mK with a sample containing a 90 nm wide quantum well on the p-side of the barrier. A heterostructure of the type of Fig. SM.1(b) in the supplement is used here. Note that no magnetic field is applied. Even without any magnetic field, large and regular oscillations of both the capacitance and the conductance are observed. The capacitance oscillations begin at about 1.7 V and are observed up to 1.9 V where the measurement has been stopped. The peak values exceed 700 $\text{nF}/\text{cm}^2$ in this sample. The geometric capacitance  would be about 500  $\text{nF}/\text{cm}^2$ using a 2DEG/2DHG distance of 21 nm derived from \textit{Nextnano} simulations.
  The periodicity of the oscillations in this sample is very regular with a period of 20.9 mV with a linear regression error of $ \pm $0.15 mV. 

The periods of the oscillations are surprisingly regular with no significant deviation from the average. The ohmic conductance shows also a periodicity but with a different shape. The regularity of the oscillations rules out that they are caused by tunneling across the X-band valley, which was found to produce current anomalies in the tunneling through much thinner barriers in similar structures.\cite{Finley1998}

\begin{figure}
	\centering
	\includegraphics[width=\breite\columnwidth]{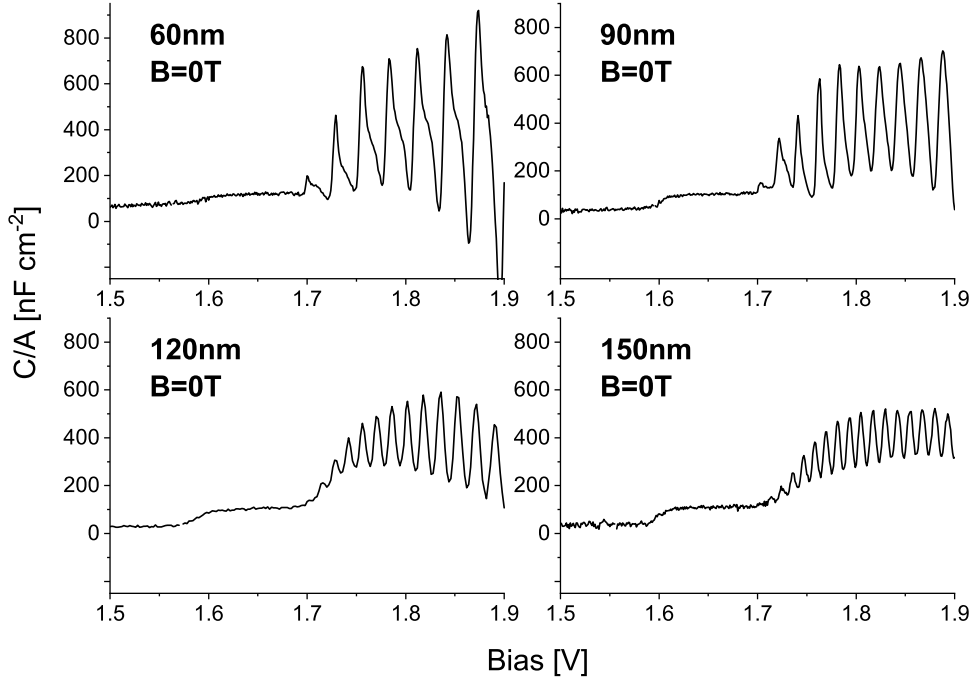}
	\caption{{\small \textit{Capacitance oscillations in four different samples with a 10-nm barrier containing quantum wells with different thicknesses. Data are taken at 4 K. Both the periods of the capacitance oscillations and their amplitudes increase with the inverse quantum well thicknesses.}}}
	\label{fig:CapDiffWells}
\end{figure}

Also noteworthy is that the measured capacitance below 1.7 V is much smaller than expected from the geometry and that the onset voltage has shifted to 1.6 V, which is a larger value than the 1.53 V for the 20 nm barrier sample of  Fig.~\ref{fig:20nmLandauOscillations}. This shift cannot be explained by the bandstructure calculations based upon \textit{Nextnano}. It occurs in all 10 nm barrier samples with the quantum well on the p-side but not in those with the 20 nm barrier that are otherwise identical. 

Returning to the oscillations above 1.7 V, we test if the oscillations are related to resonant tunneling into the quantum well. For this purpose a series of wafers with well thicknesses of  60 nm, 90 nm, 120 nm, and 150 nm, respectively, are grown. This leads indeed to dramatic changes of both the oscillation periods and their amplitudes. This is demonstrated in the data presented in Fig.~\ref{fig:CapDiffWells} where capacitance oscillations of the four samples are shown.  Oscillations are seen at all well thicknesses studied, but their periods vary. The periodicities are compiled in Table~\ref{tab:barriercompilation}.

\begin{table}[h]
\begin{tabular}{l c c c}%|}
	\toprule
QW & Exp. Period & C amp. &Theo. Period \\[-0.3cm]
(nm) 		& (mV)			& ($\text{nF}/\text{cm}^2$) & (mV)\\
	\hline
60 & 29.4$ \pm $0.3  &620  & 34.9$ \pm $0.5\\
	%\hline
90  & 20.9$ \pm $0.1 &460  & 25.9$ \pm $0.4\\
	%\hline
120 & 15.8$ \pm $0.3 &300  & 19.6$ \pm $0.2\\
	%\hline
150 & 12.0$ \pm $0.1 &220  & 16.7$ \pm $0.2\\
\toprule
\end{tabular}	
\centering\caption{\textit{{\small Data of four samples with four different quantum well thicknesses show that both the periods and the amplitudes of the capacitance oscillations increase with decreasing quantum well thicknesses. Theoretical periods are obtained from calculating the coincidences between electron levels on the two sides of the barrier as function of the bias voltage. The statistical errors of determining the respective periods are also given.}}}
\label{tab:barriercompilation}
\end{table}

The voltage periods scale inversely with the well thickness and can be numerically described by the relation $ \Delta{\text{V}}=1.34 \text{V}/(d/\text{nm}) $. Interestingly, the amplitudes of  the capacitance oscillations also scale roughly with the inverse of the well thickness. Amplitudes at the arbitrarily chosen bias voltage of 1.8 V are also given in Table~\ref{tab:barriercompilation}.

  \begin{figure}
	\centering
	\includegraphics[width=\breite\columnwidth]{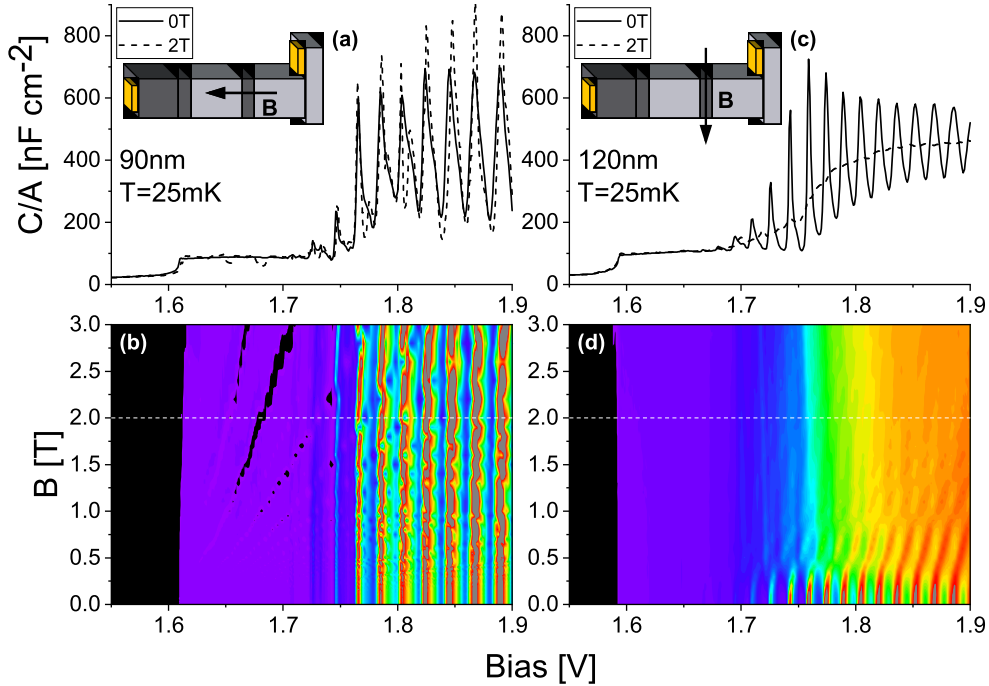}
	\caption{\textit{{\small The capacitance oscillations do not change significantly in a perpendicular magnetic field ((a) and (b)). A magnetic field oriented parallel to the barrier reduces the oscillations substantially already at 0.5 T ((c) and (d)).}}}
		\label{fig:B-field}
\end{figure}

 The oscillations are robust and change little with temperature. For example, there is little difference between 25 mK (Fig.~\ref{fig:10nmZeroFieldOscillations}) and 4 K (Fig.~\ref{fig:CapDiffWells}) for the 90 nm sample. We observe them from the mK range up to temperatures of almost 80 K. At temperatures above about 10 K they start to shift in position because the GaAs band gap changes. Conductivity oscillations have also been observed in DC experiments; an example will be shown later in Fig.~\ref{fig:Elektrolumineszenz}.

In some test samples we placed the n-doping also in AlGaAs, as on the p-side, which corresponds to the design shown in Fig. SM.1(a) of the supplement. With the doping in AlGaAs on both sides, one gets wide quantum wells on both sides of the barrier. Capacitance oscillations are observed with similar periodicities and amplitudes as shown in Fig.s~\ref{fig:10nmZeroFieldOscillations} to ~\ref{fig:B-field}. No zero field capacity oscillation are observed with the design of Fig. SM.1(c), which does not contain a quantum well. See Fig. SM.2 from the supplement.\cite{Sup} This series demonstrates that the quantum well on the p-side is the main prerequisite for the observation of the capacitance oscillations.

\textit{\textbf{Magnetic-field effect}}. A perpendicular magnetic field has a small effect on the oscillations as can be seen in Fig.~\ref{fig:B-field}(a) and (b), showing the capacitance vs. voltage as function of the B-field in a color plot for a 90 nm sample. The oscillations remain essentially unchanged with magnetic field up to 10 T exceeding the range shown in the figure. Only small irregularities in the oscillations are visible. Faint Landau oscillations are discernible between 1.6 and 1.7 V, but do not match the behavior expected for the 2DEG in this structure. It is unclear why no clear Landau oscillations are seen in the voltage range before the oscillations begin. 

In contrast, a magnetic field oriented parallel to the layers leads to a reduction of the oscillation amplitude already at fields of about 0.5 T as shown in Fig.~\ref{fig:B-field}(c) and (d). This indicates that the origin of the oscillations is connected with the electron momentum perpendicular to the interface. Assuming an electron energy of 3 meV, which would be a typical Fermi energy of a 2DEG, one expects a cyclotron radius of 90 nm. This value is large compared to the barrier thickness but comparable to the well thickness of 120 nm. This supports also that the capacitance oscillations are directly connected to the quantum well on the p-side of the barrier.

\section{Discussion} 

\begin{figure}
	\centering
	\includegraphics[width=\breite\columnwidth]{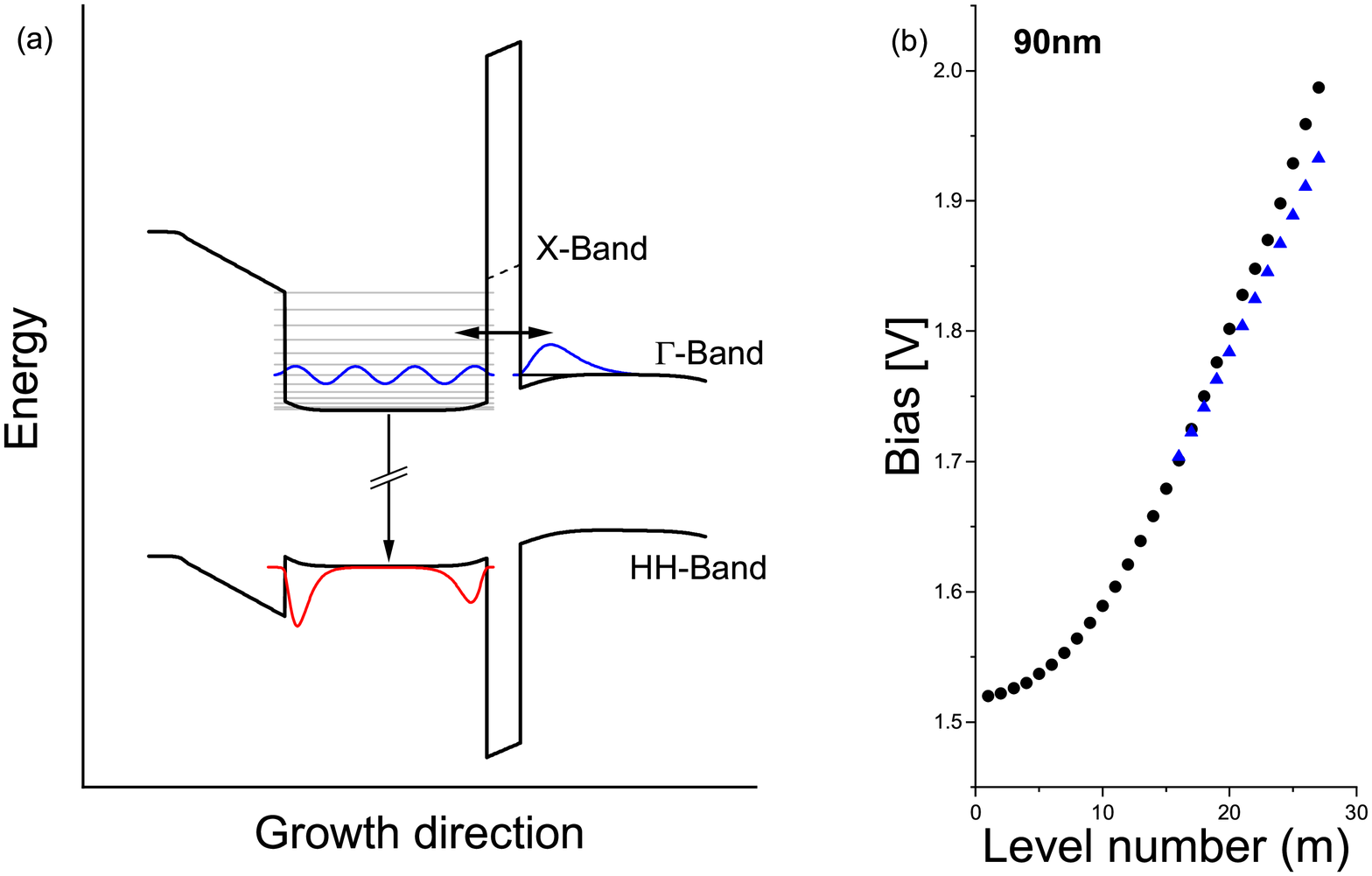}
	\caption{\textit{{\small Model for resonant tunneling of electrons from the  ($ E_0 ^n $) band into one of the  ($ E_m ^p $) bands of the GaAs quantum well on the p-side of the barrier. On the right, the voltages where simulations find a coincidence of  ($ E_0 ^n $) with  ($ E_m ^p $) are plotted as function of level number "m", black dots. The voltages of the measured capacitance maxima are contained in the same figure, shown as blue triangles. The beginning of the level number sequence is used as a fit parameter.}}}
	\label{fig:Model}
\end{figure}

To model our experimental results, we assume that the oscillations at the 10 nm barrier samples originate from resonant tunneling of electrons from the ($ E_0 ^n $) band on the n-side into unoccupied high energy levels (index "m") of the quantum well on the p-side  ($ E_m ^p $). Tunneling of the holes can be disregarded because of their larger effective mass.
Resonant tunneling into higher levels of a quantum well has been observed before \cite{Rainer1995}. However, the equal spacing of the capacitance peaks makes it at first unlikely because the quantum wells in our structures appear from Fig.~\ref{fig:LayersAndBandstructures} rather square-shaped  than parabolic. On the other hand the two 2DHGs in the valence band deform the conduction band towards parabolicity and the interplay of the voltage dependence of the energy levels on two sides has also to be considered. We use the simulation program \textit{Nextnano} to calculate the electron energy levels both on the p- and the n-side at voltages exceeding the GaAs band gap energy. Fig.~\ref{fig:Model}(a) shows the bandstructure as well as the electron levels on both sides of the barrier. Tunneling is possible if the occupied band on the n-side  ($ E_0 ^n $) lines up with one of the unoccupied electron bands  ($ E_m ^p $) on the p-side.
The voltage values where the ($ E_0 ^n $) coincide with an ($ E_m ^p $) are plotted in Fig.~\ref{fig:Model}(b) vs. the quantum number "m". 

The plot shows a surprisingly linear behavior in the voltage range comparable with the one of the measured capacitance oscillations. For the experimentally relevant voltage range (1.7 to 1.9 V) a linear fit of the values of Fig.~\ref{fig:Model}(b) finds a theoretical voltage period of 25.7 mV, which is only slightly larger than the experimentally observed 20.9 mV. The simulated level coincidences for the other quantum well thicknesses show also a nearly linear behavior in this range and the theoretical periodicities together with the statistical errors are added to the Table~\ref{tab:barriercompilation}. The values from the simulation agree reasonably well with the observed ones but are 10 to 30\% larger than the experimental ones. 

The magnetic field results fit into the resonant tunneling picture. The perpendicular magnetic field leads to Landau levels on both sides of the barrier and tunneling will be energetically possible at the same bias voltages as in zero magnetic field and there is no significant change of the oscillations. In a magnetic field parallel to the layers, the wave functions of the high index levels will be modified by the Lorentz force. This would severely reduce resonant tunneling because the in-plane momentum is no longer conserved.\cite{EisensteinGramila1991}

%The large magnitude of the capacitance oscillations remains puzzling. 
There is no obvious reason why tunneling, resonant or not, should lead to an enhanced capacitance. The capacitance should be determined by the distance between the majority carriers at the two sides of the barrier, the 2DHG and the 2DEG, respectively. Considering the finite extent of the respective wave functions it appears unlikely that their relative distance could be reduced below the 21 nm obtained from the \textit{Nextnano} simulations, which corresponds to a geometrical capacitance of about 500 $\text{nF}/\text{cm}^2$ at most. Incidentally, this capacitance value is approached in a large in-plane magnetic field in Fig.~\ref{fig:B-field}(c).

Quantum capacitance effects cannot enhance a capacitance unless the density of states turns negative. 
Indeed, a negative density of states has been invoked in theories \cite{Skinner2010} to explain anomalously large capacitance values at small densities  \cite{Li2011}. They refer, however, to the capacitance across a barrier. In our case, the origin of the enhanced capacity seems to be the quantum well at the barrier.
%Indeed, a negative density of states has been invoked in theories \cite{Skinner2010} to explain anomalous capacitance values at small densities \cite{Li2011}, but these are probably not directly applicable to our experimental situation.

Instead, we speculate that the relaxation and recombination of the electrons injected into the high-index levels of the quantum well is slow, even slower than the tunneling lifetime. Thus, after tunneling, the effective  Fermi energy of the electrons is close to that on the n-side. Thus, the electrons may tunnel back rather than relax to the bottom of the quantum well. In this case, the capacitance peaks would reflect a smaller effective distance between electrons and holes in the quantum well. 

 The anticipated long lifetime of electrons in the high levels of the quantum well is reminiscent of dark excitons, which have a long lifetime because spin-mismatch does not allow radiative recombination.\cite{PhysRevLett.126.067404} Here, it would be the small overlap of the wave functions of the holes and electrons in the quantum well that leads to long lifetimes. Long lifetimes of electrons and holes have also been achieved by spatially separating them in a lateral electric field.\cite{Zimmermann1292}
 
It is not clear why the relaxation by the emission of optical or acoustical phonons does not appear to be effective enough to relax the electrons after the resonant tunneling process towards the band edge. 
 
%Other mechanisms for fast electron relaxation can also ruled out. Transitions between the levels towards the band edge by emission of optical phonons are not possible because the energy difference is less than the one (about 36 meV) required for optical-phonon emission. Interband relaxation via acoustic phonons or electron-electron scattering tend to become very slow at low temperatures and small electron concentrations.\cite{Smet1996} Thus, only a small fraction of the injected electrons relaxes to lower levels and recombine with the holes. 

  \begin{figure}
	\centering
	\includegraphics[width=\breite\columnwidth]{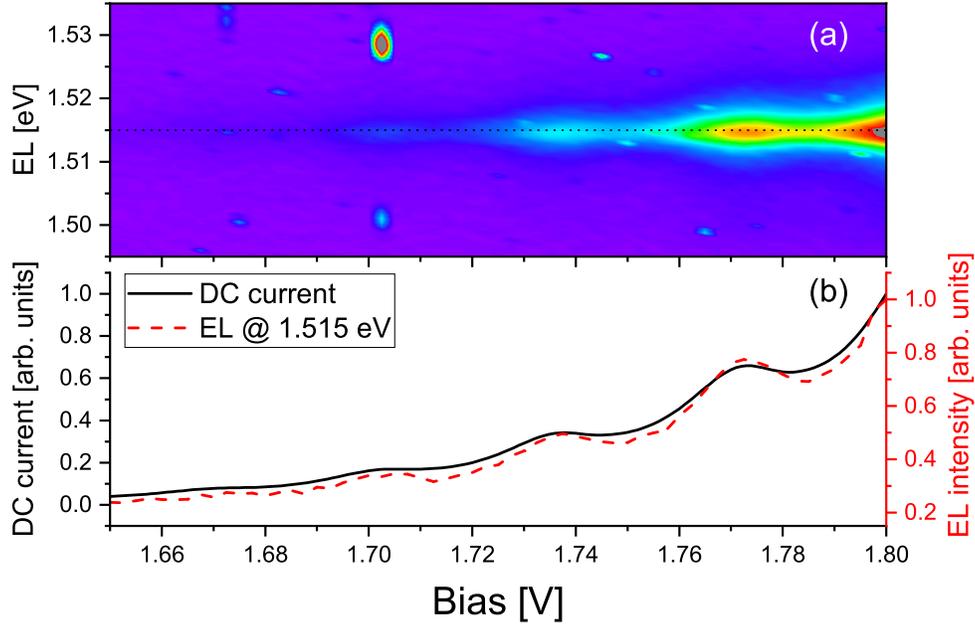}
	\caption{\textit{{\small (a) Spectral distribution of the electro-luminescence (EL) of a 10-nm barrier sample in zero magnetic field. Luminescence is observed with a photon energy of about 1.515 eV corresponding to the band gap energy of GaAs as soon as an ohmic conductance is observed. Direct recombination should emit luminescence at higher energies but is not observed. (b) Luminescence intensity at the maximum of the emission (1.515 eV) show the same oscillations as function of the voltage as the total ohmic current across the device. }}
	}
	\label{fig:Elektrolumineszenz}
\end{figure}

%Within this model, no luminescence from electrons in the high energy levels recombining with holes is expected.
 We measure the electro-luminescence and find no luminescence with photon energies higher than the one expected from the GaAs-band gap. 
 In the supplement (Fig. SM.5) the spectrally resolved electro-luminescence is shown for photon energies between 1.27 eV and 1.88 eV at bias voltages between 1.65 and 1.8 V.  Only luminescence with photon energies corresponding to the GaAs band is observed. That part of the spectra is shown in more detail in Fig.~\ref{fig:Elektrolumineszenz}(a). It shows the spectrally resolved luminescence as function of the bias voltage. Luminescence with a photon energy of 1.515 eV and a halfwidth of about 5 meV is observed at biases above about 1.65 V. This photon energy corresponds to transitions between the lowest electron-energy state with the highest hole state.
In Fig.~\ref{fig:Elektrolumineszenz}(b)  the intensity of the luminescence at 1.515 eV and the total DC conductance across the device is shown.
 The oscillations in luminescence clearly coincide with oscillations in the total ohmic conductance. Note, that the oscillations are less pronounced than in the other figures, which show differential signals. We note that the luminescence is rather weak and required high sensitivity, which is the origin of the spurious signals in Fig.~\ref{fig:Elektrolumineszenz}(a).

The origin of the very small capacitance values at voltages between the threshold and 1.7 V in Fig.s~\ref{fig:10nmZeroFieldOscillations}, \ref{fig:CapDiffWells}, and \ref{fig:B-field} is not clear. It may hint towards phenomena like exciton crystals or a BEC that are conceivable in the small density regime.\cite{Skinner2016} It may also be that the charge distribution is different from the one expected from the simulations. Presently, we do not have any information if any of these phenomena play a role for the observations at small densities.

 \section{Conclusions}
%Using MBE we have produced p-i-n structures with a 10-nm barrier and apply a forward bias to induce 2DEG/2DHG bilayers separated by a few nm. Large peaks in the capacitance values are observed at very regular voltage periods. We suggest that these peaks are caused by resonant injection of electrons into the high energy levels of a wide quantum well that exists on the p-side of the device. The long life times of the electrons in these levels cause the large capacitance values. This could indicate the existence of long-lived excitons in the well. In contrast, at voltages near the  GaAs band gap energy, the capacitances are smaller than expected.This phenomenon could be caused by the Coulomb interaction between the layers.         
At the outset of this research, our goal was to find a BEC of excitons from the 2DEG/2DHG bilayers separated by a few nm. 
There is no indication in our results of a BEC of excitons consisting of the opposite charges on the two sides of the barrier.
Instead we find that our structures host an unusual and possibly novel interacting charge system. It resides in a quantum well that contains holes intrinsically while electrons are tunnel-injected selectively into high energy levels. The large peak capacitance values could be caused by a negative density of states, which would point to a correlated charge system. The microscopic nature of this charge system cannot be determined by our data. It might be liquid-like, have a charge order, or form a condensate.

 \section{Acknowledgement}  This work profited from preliminary studies by Michael Pohlt and Mark Lynass during their PhD work at the MPI für Festkörperforschung, Stuttgart and from Stefan Fälts suggestion to put the doping into an AlGaAs layer. We appreciate discussions with Jérôme Faist, Thomas Ihn, Andisheh Khedri, Klaus von Klitzing, Jörg Kotthaus, Antonio Štrkalj, and Oded Zilberberg. Furthermore we acknowledge the financial support of the Swiss National Foundation (SNF) and the NCCR QSIT (National Center of Competence in Research - Quantum Science and Technology).

\bibliography{BibFiles/Fused_verbessert}

\end{document}